# Enhancing Generalization in PPG-Based Emotion Measurement with a CNN-TCN-LSTM Model

Karim Alghoul
School of Electrical Engineering and
Computer Science
University of Ottawa
Ottawa, Canada
Karim.Alghoul@uottawa.ca

Hussein Al Osman
School of Electrical Engineering and
Computer Science
University of Ottawa
Ottawa, Canada
Hussein.Alosman@uottawa.ca

Abdulmotaleb El Saddik
School of Electrical Engineering and
Computer Science
University of Ottawa
Ottawa, Canada
elsaddik@uottawa.ca

*Abstract*—Human-computer interaction has become integral to modern life, driven by advancements in machine learning technologies. Affective computing, in particular, has focused on systems that recognize, interpret, and respond to human emotions, often using wearable devices, which provide continuous data streams of physiological signals. Among various physiological signals, the photoplethysmogram (PPG) has gained prominence due to its ease of acquisition from widely available devices. However, the generalization of PPG-based emotion recognition models across individuals remains an unresolved challenge. This paper introduces a novel hybrid architecture that combines Convolutional Neural Networks (CNNs), Long Short-Term Memory networks (LSTMs), and Temporal Convolutional Networks (TCNs) to address this issue. The proposed model integrates the strengths of these architectures to improve robustness and generalization. Raw PPG signals are fed into the CNN for feature extraction. These features are processed separately by LSTM and TCN. The outputs from these components are concatenated to generate a final feature representation, which serves as the input for classifying valence and arousal, the primary dimensions of emotion. Experiments using the Photoplethysmogram Dataset for Emotional Analysis (PPGE) demonstrate that the proposed hybrid model achieves better model generalization than standalone CNN and LSTM architectures. Our results show that the proposed solution outperforms the state-of-the-art CNN architecture, as well as a CNN-LSTM model, in emotion recognition tasks with PPG signals. Using metrics such as Area Under the Curve (AUC) and F1 Score, we highlight the model's effectiveness in handling subject variability.

I. INTRODUCTION

Advances in signal analysis for speech and facial expressions have been pivotal in the field of affective computing [1],[2]. Machine learning has significantly contributed to this progress, with algorithms developed to classify emotional states from inputs such as speech, facial expressions, and gestures [3],[4],[5]. Among these techniques, neural networks, in particular, enable automatic emotional feature extraction, improving recognition accuracy [6].

Research studies have proposed various potential applications for these advancements. For instance, in AI-based education, adaptive systems enhance learning by responding to students' emotional states [7]. Similarly, in healthcare, emotion-aware machines can assist medical professionals in identifying distressed patients who require immediate intervention [7]. These advancements are further enhanced by the integration of wearable devices and smart clothing, which generate continuous and personalized data streams.

The introduction of wearable technology has enabled the non-invasive, continuous acquisition of physiological signals, such as electroencephalograms (EEG), electrocardiograms (ECG), and photoplethysmograms (PPG) [8]. These signals provide objective and reliable indicators of emotional states, as they are less susceptible to social masking compared to physical signals like facial expressions or voice tones [9],[10]. The widespread availability of wearable technology has also driven the development of advanced machine learning algorithms capable of processing the rich data generated by these devices, enabling progress in fields such as mental health monitoring and human-computer interaction [11].

Among physiological signals, PPG offers distinct advantages for emotion recognition, particularly when integrated into smartwatches. Unlike EEG or ECG sensors, PPG sensors are highly portable, and unobtrusive, allowing for seamless, continuous monitoring without disrupting users' daily activities [12]. Additionally, PPG can be measured remotely using camera-based methods [13],[14],[15],[16], making it more accessible in situations where direct contact sensors may not be feasible. The ability to collect PPG signals through both wearable devices and remote imaging makes it a versatile solution for emotion recognition applications. These features, combined with the increasing affordability and accessibility of wearable devices, have popularized emotion recognition technologies, making them more practical for a broader audience [17].

Despite these advancements, the generalization of emotion recognition models remains a significant challenge, particularly for PPG-based systems. Many models achieve high accuracy in controlled settings but fail across diverse subjects, limiting real-world applicability. Generalization refers to a model's ability to perform well on unseen data [18]. Metrics such as test accuracy alone is insufficient; more comprehensive evaluations, such as AUC and F1 score, better assess robustness. AUC, in particular, measures class separation across decision thresholds, providing deeper insight into model performance beyond training data [19], [20], [21]. The importance of the AUC metric is discussed in more details in section IV.B. Neglecting these metrics risks developing models that excel in controlled settings but fail to generalize in real-world scenarios.

**Objectives and Contributions**

This paper seeks to address the gap in model generalization for emotion recognition from PPG signals. To this end, we propose a novel hybrid model that combines the strengths of CNNs, LSTMs, and TCNs. The proposed model is designed to extract robust feature representations and enhance model generalization in emotion detection tasks.

The key contributions of this work are:

1. We introduce a hybrid architecture that uniquely combines CNN, LSTM and TCN components, leveraging their complementary strengths for PPG-based emotion recognition.
2. We evaluate our model on the PPGE dataset [22] using the LOSO cross-validation approach, which is particularly suited for assessing cross-subject variability, and compare its performance against the state-of-the-art CNN architecture and a CNN-LSTM model. The evaluation is performed using AUC, accuracy, and F1 score to assess classification performance and generalization ability.

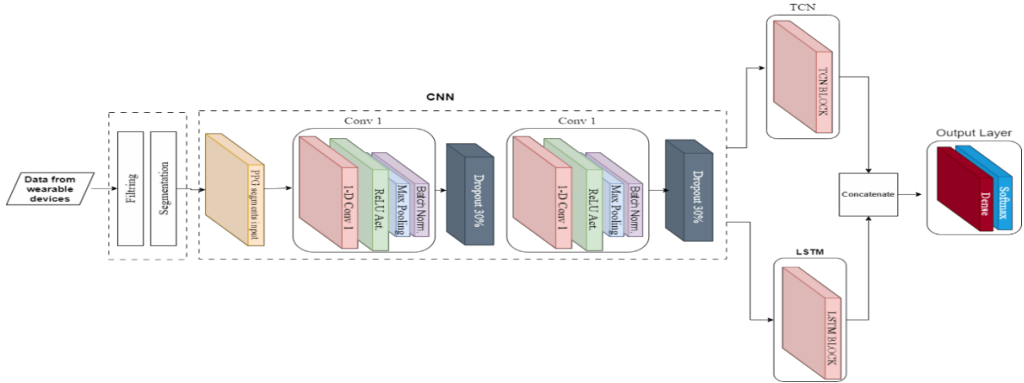

Figure 1 Overview of the Proposed Model

## II. RELATED WORK

While PPG is commonly integrated into multimodal Emotion Recognition Systems (ERS) [23], [24], [25], [26], [27], its standalone application remains limited due to sensitivity to noise. However, recent studies suggest that PPG, despite being less frequently utilized as a unimodal input for emotion recognition [28], can achieve results comparable to ECG when used as the sole input for ERS [28],[29],[30],[31].

The advantages of PPG over other physiological signals make it a particularly attractive option for real-world applications. It can be easily obtained using wearable devices like smartwatches and smartphones, eliminating the need for bulky EEG headsets or ECG belts. Wrist and finger-based PPG sensors are user-friendly, cost-effective, and non-invasive, making them ideal for continuous monitoring [29]. These factors have increased interest in its standalone use for emotion recognition.

Recent studies have proposed various architectures for PPG-based emotion detection. For example, Lee M. et al. [29] used a 1D CNN for short-term emotion recognition based on a single pulse PPG signal, validated on the DEAP dataset [32]. Similarly, [30] proposed parallel CNN architectures processing normalized PPG segments and NN intervals. Additionally, [31] introduced a hybrid CNN (H-CNN) for stress detection, combining hand-crafted and automatically learned features, validated on the WESAD dataset [33].

Beyond CNNs, Long Short-Term Memory (LSTM) networks and hybrid CNN-LSTM architectures have also been explored for emotion recognition. For instance, Etienne et al. [34] developed a CNN-LSTM model for emotion recognition from speech data, while [35] demonstrated the application of LSTMs in physiological signal processing. Pre-trained CNNs combined with residual LSTMs have been applied to EEG signals for emotion detection [36]. Regarding PPG signals, CNN-LSTM architectures have been effectively used, as shown in [37], which employed a hybrid model combining 1D-CNN and LSTM for emotion recognition tasks using remote PPG signals.

Temporal Convolutional Networks (TCNs) have emerged as another promising architecture ensuring no future information leaks while maintaining the same output size as input [38]. Harb [39] demonstrated the feasibility of TCNs for emotion recognition through a multimodal model incorporating audio, visual, and textual features. TCNs have also been applied to EEG and ECG-based emotion recognition [40], [41], [42].

For PPG signals, while TCNs have been used in heart rate monitoring [43] and engagement classification in E-learning [44], their application for emotion recognition remains unexplored. To our knowledge, no prior work has combined TCNs with PPG for this task.

## III. PROPOSED METHODOLOGY

### A. Proposed architecture

Our proposed model architecture combines Convolutional Neural Networks (CNNs), Long Short-Term Memory networks (LSTMs), and Temporal Convolutional Networks (TCNs) to leverage the strengths of each network type for enhanced feature extraction and temporal modeling in physiological signal analysis. Each of these components is selected based on its unique ability to process specific aspects of PPG signals. This hybrid approach addresses the limitations of standalone models by integrating complementary capabilities. The architecture is illustrated in Figure 1.

The model begins with a shared CNN branch that processes the input signal through a series of convolutional layers. CNNs have been widely used for PPG-based emotion recognition due to their ability to extract meaningful spatial features from physiological signals. Previous studies have shown that CNNs can effectively capture morphological characteristics of PPG waveforms, such as pulse amplitude and waveform shape, which are indicative of emotional states [29],[30]. In particular, 1D CNNs have been successfully applied for short-term emotion recognition [29] and stress detection [31]. Given their demonstrated success, we employ a CNN-based feature extractor as the first stage of our model.

The first CNN layer applies 8 filters with a kernel size of 64 and a stride of 4, using *ReLU* activation and *'same'* padding. This is followed by *max pooling* and *batch normalization* to reduce dimensionality and stabilize learning, respectively. To prevent overfitting, a *dropout* layer with a 30% rate is included. A second convolutional layer, comprising 16 filters with a kernel size of 32 and a stride of 2, further refines the feature maps. This layer is also followed by *max pooling*, *batch normalization*, and *dropout* to ensure robust feature extraction.

The processed features from the CNN layers are then fed into two parallel branches: a TCN branch and an LSTM branch. PPG signals exhibit temporal variations that reflect physiological responses to emotional stimuli. LSTMs are well-suited for capturing such sequential dependencies, as they have been successfully applied to physiological signal processing tasks, including emotion recognition [31],[33]. Previous works have leveraged hybrid CNN-LSTM models to extract spatial and temporal features from PPG signals, demonstrating improved classification performance compared to CNN alone [33]. Motivated by these findings, we integrate an LSTM branch to enhance our model's ability to capture long-term temporal dependencies. Similarly, TCNs have been applied to various emotion recognition tasks involving EEG and ECG signals [36],[37],[38], but their use in PPG-based emotion recognition remains largely unexplored. Inspired by their effectiveness in other physiological signal domains, we incorporate a TCN branch to capture hierarchical temporal patterns in PPG signals.

**TCN Branch:** This branch employs a Temporal Convolutional Network (TCN) with 8 filters, a kernel size of 32, and dilation rates set to [1, 2, 4, 8]. Causal padding is used to maintain the temporal order of the input signal. Skip connections are incorporated to improve gradient flow, and a *dropout* rate of 30% is applied to mitigate overfitting. We implemented the TCN using the KerasTCN library [45].

**LSTM Branch:** This branch includes an LSTM layer with 12 units to capture sequential dependencies and long-range temporal dynamics within the data. The LSTM focuses on modeling the inherent temporal patterns in the signal.

The outputs from the TCN and LSTM branches are concatenated to create a combined feature representation that integrates spatial and temporal information. This fused representation is passed to a fully connected output layer consisting of two neurons with a *SoftMax* activation function, enabling the model to perform binary classification for valence or arousal states.

While CNNs focus on extracting spatial features, they lack the ability to model sequential dependencies. LSTMs capture these dependencies but struggle with hierarchical temporal structures [46]. TCNs efficiently handle long-range dependencies but do not inherently extract spatial features. By integrating all three architectures, we create a comprehensive feature representation that leverages their complementary strengths. By combining the spatial feature extraction capability of CNNs, the sequential modeling strengths of LSTMs, and the hierarchical temporal processing power of TCNs, our model addresses key challenges in PPG-based emotion recognition, improving robustness and generalization compared to standalone CNN and LSTM models.

*B. Data Preprocessing*

There are two primary approaches for feature extraction from physiological measurements. The first is the hand-crafted feature method, which relies on Heart Rate Variability (HRV) analysis derived from the inter-beat intervals (IBIs). The second leverages deep learning, where features are automatically extracted using various neural network models [47],[48]. Deep learning facilitates representation learning, which reduces or eliminates the need for manual feature engineering and often results in the discovery of effective feature sets for a task [18],[12]. In this study, we focused on the second method, using raw PPG signal as the input to the model, and several preprocessing steps were undertaken to prepare the data for analysis and training.

1. **Signal Filtering**: To enhance signal quality, a third-order Butter-worth bandpass filter was applied to the raw PPG signals. The filter, designed with a cutoff frequencies of 0.7 to 3.7 Hz, captures the frequency range corresponding to heart rates from approximately 40 beats per minute (bpm) at rest to 220 bpm during exercise or tachycardia. This filter was chosen for its effectiveness in minimizing noise while preserving critical signal features. This process follows the approach described in [49].

2. **Segmentation**: The preprocessed PPG signals were segmented into smaller windows to capture temporal patterns and variations within the signal. A sliding window approach was employed, dividing the signals into segments of 60 seconds, with a 5-second overlap between consecutive windows similar, following the methodology outlined in [31]. Each segment was then assigned labels, as explained in section IV.A.

3. **Standardization**: Segments were standardized using Z-score normalization before feeding it to our model.

IV. EXPERIMENTAL EVALUATION

*A. Dataset*

Emotions can be represented using categorical or dimensional models [53]. The categorical approach assigns discrete labels (e.g., stress, joy) but lacks universal translation across languages and fails to capture the full emotional spectrum [54]. The dimensional model, particularly the valence-arousal framework, provides a continuous representation, where valence ranges from unpleasant (1) to pleasant (9) and arousal from calm (1) to excited (9) [55],[56]. We use the PPGE dataset [22] as it is the only publicly available dataset with binary emotion labels and longer video stimuli (~5 minutes), which elicit stronger emotional responses compared to shorter clips in other datasets. It contains 72 PPG signals from 18 participants (13 men, 5 women, aged 23–31), recorded at 100 Hz using a fingertip pulse sensor. After watching each video, participants rated their emotional response, forming the dataset's valence-arousal labels.

Each subject participated in four trials, with each trial corresponding to a distinct emotional stimulus. Trial information was retained to ensure that data from different trials and subjects were not mixed, preserving unique physiological responses. These factors make PPGE well-suited for validating our model. However, its main limitation is the relatively small number of participants.

### B. Performance Metrics

The accuracy metric has notable limitations, particularly its inability to measure misclassification costs, which is a significant drawback in real-world applications where these costs are often unequal [20]. This issue becomes even more pronounced in the context of imbalanced datasets. Previous studies have demonstrated the superiority of AUC over accuracy as a performance measurement, as AUC provides a more robust and informative evaluation [19],[20],[21]. Unlike accuracy, AUC offers consistency across imbalanced datasets [50], [51] by measuring sensitivity and specificity collectively over various threshold values. It represents the probability that a classifier will rank positive instances higher than negative ones, making it a more reliable measurement of performance.

This makes it particularly suitable for evaluating the model's generalization and robustness in binary classification tasks.

In addition to AUC, we still report test accuracy for an overall measure of correctness and the F1 Score (both unweighted and weighted) as it is commonly used in the literature to address and evaluate performance in imbalanced datasets. The unweighted F1 Score evaluates performance on underrepresented classes independently, while the weighted F1 Score adjusts for class frequencies.

TABLE I
LOSOCV OF OUR PROPOSED MODEL VS CNN AND CNN-LSTM MODELS ON PPGE DATASET

| Model | *Valence only* | | | | |
|---|---|---|---|---|---|
| | Test Acuuracy | F1-score class 0 | F1-score class 1 | weighted F1 | AUC |
| CNN | 0.76 | 0.40 | **0.84** | 0.62 | 0.64 |
| CNN-LSTM | 0.74 | 0.38 | 0.82 | 0.60 | 0.6 |
| Our Model | **0.77** | **0.45** | **0.84** | **0.65** | **0.66** |
| | *Arousal only* | | | | |
| CNN | 0.68 | 0.71 | 0.49 | 0.60 | 0.61 |
| CNN-LSTM | **0.73** | **0.78** | **0.55** | **0.67** | 0.68 |
| Our Model | 0.69 | 0.75 | 0.52 | 0.64 | **0.69** |
| | *Average of Valence and Arousal* | | | | |
| CNN | 0.72 | 0.56 | 0.66 | 0.61 | 0.63 |
| CNN-LSTM | **0.74** | 0.58 | **0.69** | 0.63 | 0.64 |
| Our Model | 0.73 | **0.60** | 0.68 | **0.64** | **0.68** |

### C. Model Training and Evaluation

The Leave-One-Subject-Out cross-validation (LOSOCV) method was employed for experiments involving a single dataset with a relatively small number of participants. In this approach, each participant's data is used as the test set once, while the remaining participants' data form the training set. This method evaluates the model's ability to generalize across individuals within the same dataset and provides insights into its robustness in handling subject variability.

Training Details:

- **Early Stopping:** We trained our model with a batch size of 512 for 350 epoch. Training incorporated an early stopping mechanism with a patience setting of 80 epochs. This prevented overfitting by halting training when no improvement in the validation accuracy score was observed on the validation set.

- **Weighted categorical crossentropy:** To address potential class imbalances, dynamic class weights were assigned during training. These weights ensured fair representation by emphasizing minority classes, improving the model's performance on underrepresented data.

- **Optimization and Loss Function:** The Adam optimizer, with a learning rate of 0.01 was utilized for its efficient handling of sparse gradients. The weighted categorical crossentropy was employed as the loss function.

## V. RESULTS AND DISCUSSION

CNN and CNN-LSTM architectures are widely used in the literature for emotion recognition tasks with PPG signals. Studies such as [29],[30],[31],37] have demonstrated their effectiveness in feature extraction and emotion classification from PPG signals. Based on their established use, we included CNN and CNN-LSTM as baseline models to compare against our proposed CNN-TCN-LSTM architecture.

### A. Results

Table 1 presents the performance of the models which were evaluated on the PPGE dataset for Valence, Arousal, and their average classification.

*1) Valence Classification*

Our combined model (CNN-TCN-LSTM) demonstrated the highest AUC of 0.66, indicating superior ability to generalize across subjects, and achieved the highest average F1-score of 0.65 among all models. In comparison, the CNN and CNN-LSTM models had lower AUC values of 0.64 and 0.60, respectively, reflecting weaker generalization than the combined model. This highlights the effectiveness of integrating CNN, LSTM and TCN components for robust feature extraction and temporal modeling.

*2) Arousal Classification*

Our combined model again achieved the highest AUC of 0.69. In comparison, while the CNN-LSTM model performed better in terms of average F1-score (0.67), its lower AUC of 0.68 reflects a reduction in model generalization. Meanwhile, the CNN model lagged significantly, with an AUC of 0.61 and the lowest average F1-score of 0.60.

*3) Average Valence and Arousal*

When examining the average values of both Valence and Arousal, it is evident that our combined model outperforms the other two models in both AUC and F1-score. The combined model achieves an AUC of 0.68 and an average F1-score of 0.64, compared to AUC values of 0.64 and 0.63 and F1-scores of 0.63 and 0.61 for the CNN-LSTM and CNN models, respectively. These results confirm that the hybrid architecture offers a significant improvement in generalization.

*B. Discussion*

Our combined model remains the best-performing architecture, even though the CNN-LSTM model achieved a higher test accuracy of 0.74. However, test accuracy alone is insufficient for evaluating model generalization [52],[53]. This is why AUC is prioritized as the primary metric, as it is a more reliable indicator of model generalization, particularly for imbalanced datasets and binary classification problems [19],[20]. The combined model consistently demonstrated the highest AUC across Valence, Arousal, and their combined classification tasks, highlighting its superior ability to generalize across subjects.

These results reinforce PPG's effectiveness in capturing emotional states, highlighting its advantages as a non-invasive, easily accessible signal from wearable and camera-based devices. Unlike EEG or ECG, which require specialized sensors, PPG enables seamless, continuous monitoring in real-world settings. While PPG alone performs well, its accuracy can improve through multimodal integration. Combining PPG with facial expressions or voice intonation could enhance robustness, especially in noisy environments. Future research should explore such fusion to develop more generalized emotion recognition models.

## VI. CONCLUSION

In this paper, we introduced a novel CNN-TCN-LSTM combined model for emotion recognition from PPG signals, addressing critical challenges in affective computing, particularly the model generalization of emotion recognition models. The proposed model integrates the complementary strengths of CNNs for spatial feature extraction, LSTMs for modeling sequential dependencies, and TCNs for capturing hierarchical temporal patterns. To evaluate the model, we utilized the PPGE dataset and conducted experiments using Leave-One-Subject-Out (LOSO) evaluation approach, which is well-suited for assessing cross-subject generalizability.

The results demonstrated that our combined model consistently outperformed CNN and CNN-LSTM standalone architectures, achieving the highest AUC across Valence, Arousal, and their average classification tasks. AUC was prioritized as the primary metric, as it provides a more reliable indicator of model generalization, particularly in binary classification settings and imbalanced datasets. While the CNN-LSTM model showed competitive validation accuracy and F1-scores for specific classes, our combined model's superior AUC values underscore its robustness and stability across different subjects.

In future work, we aim to investigate the model's performance in multi-class emotion recognition tasks and evaluate its effectiveness in real-world scenarios using wearable devices with varying signal quality. Additionally, we plan to extend this study by exploring the use of multi-modal approaches that incorporate additional physiological signals, such as ECG and EEG, to further enhance the model's classification accuracy and model generalization.